\documentclass[journal=jctcce,manuscript=article]{achemso}

\usepackage[version=3]{mhchem} 

\usepackage{array}
\usepackage{indentfirst}
\usepackage{arydshln}
\usepackage[english]{babel}
\usepackage{graphicx}
\usepackage{subfig}
\usepackage{amsmath}
\usepackage{algorithm}
\usepackage{algorithmic}
\usepackage{moresize}
\usepackage[labelfont=bf]{caption}
\usepackage{amssymb}
\usepackage[svgnames]{xcolor}
\usepackage{epsfig,euscript}
\usepackage{color}
\usepackage{geometry}
\DeclareTextSymbol{\degre}{OT1}{23}

\usepackage{pgfplots}                                           
\usepackage{pgf}
\usepackage{tikz}                                               
\usepackage{multirow}

\author{Bastien Casier}
\affiliation{Universit\'e de Lorraine and CNRS, LPCT, UMR 7019, F-54000 Nancy, France}
\email{bastien.casier@univ-lorraine.fr}

\author{Mauricio Chagas da Silva}
\affiliation{Universit\'e de Lorraine and CNRS, LPCT, UMR 7019, F-54000 Nancy, France}

\author{Michael Badawi}
\affiliation{Universit\'e de Lorraine and CNRS, LPCT, UMR 7019, F-54000 Nancy, France}

\author{Fabien Pascale}
\affiliation{Universit\'e de Lorraine and CNRS, LPCT, UMR 7019, F-54000 Nancy, France}

\author{Tom\'{a}\v{s} Bu\v{c}ko}
\affiliation{Department of Physical and Theoretical Chemistry, Faculty of Natural Sciences, Comenius University in Bratislava, Mlynsk\'{a} Dolina, Ilkovi\v{c}ova 6, SK-84215 Bratislava, Slovakia}
\altaffiliation{Institute of Inorganic Chemistry, Slovak Academy of Sciences, D\'{u}bravsk\'{a} cesta 9, SK-84236 Bratislava, Slovakia}

\author{S\'ebastien Leb\`egue}
\affiliation{Universit\'e de Lorraine and CNRS, LPCT, UMR 7019, F-54000 Nancy, France}

\author{Dario Rocca}
\affiliation{Universit\'e de Lorraine and CNRS, LPCT, UMR 7019, F-54000 Nancy, France}
\email{dario.rocca@univ-lorraine.fr}

\title{Hybrid localized graph kernel for machine learning energy-related properties of molecules and solids}

\date{\today}

\begin{document}

\begin{abstract}

Nowadays, the coupling of electronic structure and machine learning techniques serves as a powerful tool to predict chemical and physical properties of a broad range of systems. With the aim of improving the accuracy of predictions, a large number of representations for molecules and solids for machine learning applications has been developed. In this work we propose a novel descriptor based on the notion of molecular graph.  While graphs are largely employed in classification problems in cheminformatics or bioinformatics, they are not often used in regression problem, especially of energy-related properties. Our method is based on a local decomposition of atomic environments and on the hybridization of two kernel functions: a graph kernel contribution that describes the chemical pattern and a Coulomb label contribution that encodes finer details of the local geometry. The accuracy of this new kernel method in energy predictions of molecular and condensed phase systems is demonstrated by considering the popular QM7 and BA10 datasets. These examples show that the hybrid localized graph kernel outperforms traditional approaches such as, for example, the smooth overlap of atomic positions (SOAP) and the Coulomb matrices.


\end{abstract}

\maketitle

\section{\label{sec:intro} Introduction}

The past decade has seen an impressive growth in the development and application of machine learning techniques \cite{haykin2009} in quantum chemistry and computational condensed matter physics \cite{behler2016,noe2020, dral2020}. These methods are of a great interest for theoreticians because they  allow for the analysis,  classification and  prediction of various properties conventionally requiring a large amount of data generated via computationally demanding quantum mechanical calculations \cite{mcardle2020}. 
Indeed, machine learning techniques can be applied to a broad range of problems, including, among the others, potential energy surface fitting \cite{handley2010,behler2011_2,bucko2020ab,casier2020}, \textit{ab initio} molecular dynamics \cite{behler2007,hase2019, gkeka2020}, prediction of various scalar properties \cite{montavon2013, welborn2018, pronobis2018} (\textit{e.g} atomization energies, polarizability coefficients, highest occupied molecular orbital energies, electronic structure correlation energies, \textit{etc}), and vectorial and tensorial quantities (\textit{e.g} forces, polarizability tensors, \textit{etc}) \cite{grisafi2018, unke2019}. 


The chemical compound space is characteristic by a huge dimensionality and complexity. Datasets for molecules and solids proposed in the literature, which span only a small part of the chemical space, already contain impressively large numbers of compounds. The GDB-17 molecular dataset, for instance, contains 166 billions of molecules~\cite{ruddigkeit2012}. An example of a rich dataset of crystal structures is the set of elpasolites of Faber \textit{et al.}~\cite{faber2016} containing 2 millions
crystals. This shows clearly that it is impossible to analyze or screen all these compositions and structures by demanding electronic structure calculations \cite{behler2016}. A solution to efficiently explore the chemical compound space (or at least significant parts of it) can be found in the development of new machine learning approaches to complement \emph{ab initio} calculations. This field of research is known as quantum machine learning \cite{lilienfeld2018, tkatchenko2020} (QML). 
The main idea of the QML approaches is to train a machine learning model on a subset of chemical structures (the smallest possible) for which simulations at a quantum mechanical level were done. The trained machine is then used to predict the target properties of the rest of the systems. This technique can be applied not only in the chemical composition space but also in the conformation space of a given system. For example, recently, a $\Delta$-machine learning approach \cite{ramakrishnan2015} based on thermodynamic perturbation theory has shown impressive results in the prediction of the adsorption enthalpy of small molecules adsorbed in chabazite at the random phase approximation (RPA) level of theory using only 10 training conformations \cite{chehaibou2019}. In general, the ultimate goal of QML would be to find a universal model which unifies both chemical and conformational spaces \cite{tkatchenko2020}.  

One of the main challenges in the use of QML methods is the choice of proper descriptors -- \textit{i.e.} features -- to represent the molecular and/or condensed matter systems. In this work we will focus on structural descriptors, namely approaches that assume the knowledge of the precise geometry of the systems under consideration beyond their chemical composition. Currently, a large number of those descriptors have been proposed in the literature, with the symmetry functions \cite{behler2011}, the Coulomb matrices \cite{rupp2012, montavon2012, faber2015}, the smooth overlap of atomic positions (SOAP) \cite{bartok2013, de2016} and the many-body tensor representation (MBTR) \cite{huo2017} being some of the most prominent examples. From a general point of view an efficient structural descriptor should satisfy certain prerequisites \cite{noe2020}. For instance, the descriptor should be unique (one-to-one correspondence between features and systems) and invariant with respect to overall rotations and rigid translations, as well as permutations of atoms of the same type. 

In the present work we introduce a new descriptor based on the notion of graph that satisfies all the above-mentioned conditions. Graphs are already largely used in the domain of cheminformatics \cite{mahe2004,mahe2005,gauzere2012,lavecchia2015} and bioinformatics \cite{sharan2006, smalter2009} to predict the activity or the toxicity of particular drugs  or to recognize particular patterns of binding site of biomolecules (with the quantitative structure-activity relationships (QSAR) model \cite{muratov2020}). Nevertheless, in these cases, the use of graphs is restricted to classification problems and their applications in regression problems are still rare, especially as far as energy-related properties are concerned \cite{ferre2017,wu2018,tian2018,tang2019}. These quantities are particularly sensitive to inner changes of the molecular structure which cannot be described by simple graphs whose edges are represented by 0's and 1's for disconnected and connected nodes (namely atoms)~\cite{gyoung2020}. Indeed, this simplified representation leads to isomorphic graphs even for rather different structures and this represents an issue for learning properties (\textit{e.g.} the energy) which are very sensitive on the specific geometry of a system. To overcome this difficulty in the present work we used weighted graphs, whose values on the edges were obtained as superposition of atom-centered Gaussian functions.   
In order to develop a machine approach applicable to both molecules and solids we considered graphs defined in localized environments around each atom of a given system. These localized graphs were used as input features of a
machine learning approach based on the kernel ridge regression \cite{hoerl1970} (KRR). This method scales cubically with the size of the training set but typically requires fewer data point to be trained than other traditional approaches (\textit{e.g.} neural networks). The role of the kernel in KRR is to provide a measure of similarity between different localized graphs belonging to different systems.
A global ``comparison'' of two systems (molecules or solids) can then be obtained by summing localized contributions from each atomic environment \cite{de2016}. The specific kernel used in this work combines two different parts: The first, based on a variant of the shortest-path kernel, is suited to describe structural motifs and chemical bonding between atoms; the second includes finer geometric details through a label enrichment of the nodes with Coulomb vectors. This approach can be viewed as an hybrid kernel function \cite{wu2012} and we will show that the introduction of both contributions provides a high level of accuracy. As a proof of principle this new methodology was applied to predict molecular atomization energies (QM7 dataset) and enthalpies of formations of solids (BA10 dataset). 


The paper is organized as follows. In Sec.~\ref{sec::Method}  the concept of labeled localized graph is introduced, which is represented trough maximum probability paths and enriched with a set of vectors that describe the Coulomb potential of the atomic environment. In the same section, the new kernel approach is also presented. In Sec.~\ref{sec::Benchmarks} the datasets QM7 and BA10, used to evaluate our method are reviewed. Finally, our results for the prediction of the atomization energies of molecular systems (QM7 dataset) and the formation enthalpies for the solid state systems (BA10 dataset) are presented in Sec.~\ref{sec::Results}.    

\section{\label{sec::Method} Methodology}

\subsection{\label{sec::LMG} The Labeled Localized Graph Descriptor.}
 
\begin{figure}
\centering
\includegraphics[scale=0.7]{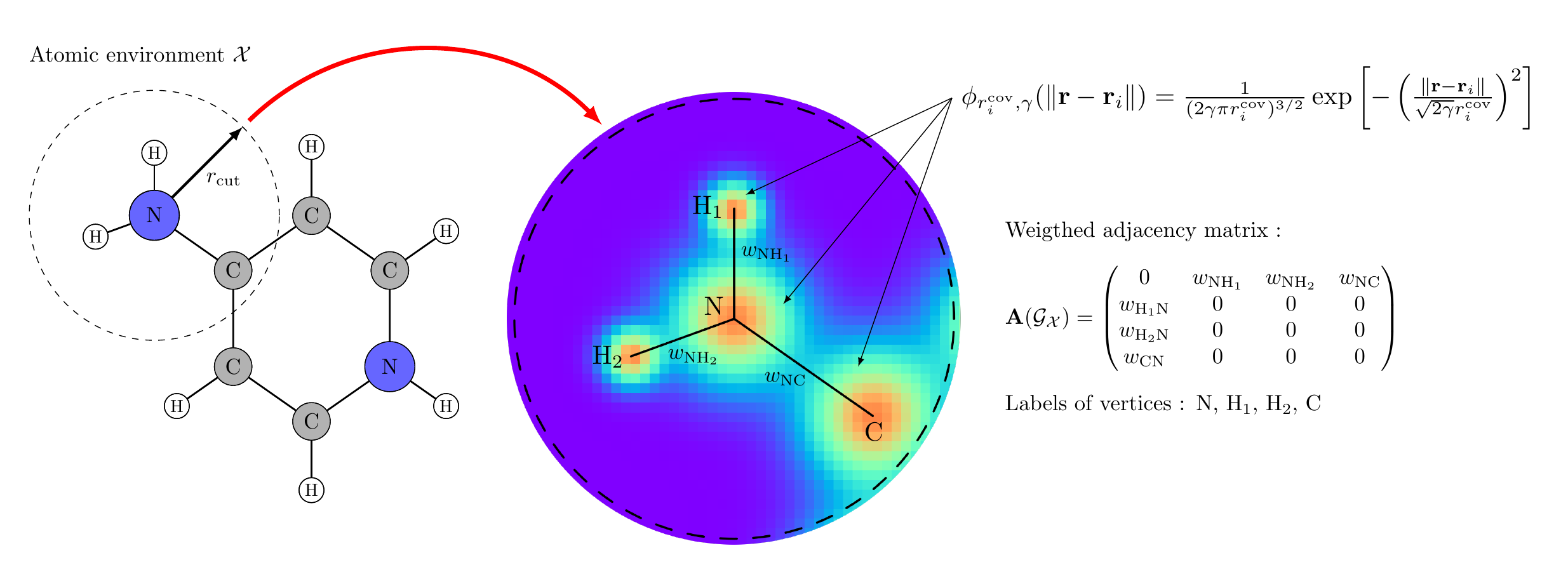}
\caption{\label{Fig::LLG} Illustration of a labeled localized graph (LLG) for a nitrogen atomic environment defined by a radius $r_{\text{cut}}$. The overlaps of the Gaussian functions associated to each atom in this environment determine the weights in the adjacency matrix of the graph.}
\end{figure}

In this section we introduce the concept of labeled localized graph (LLG) for molecules and solids. By considering a local atomic environment $\mathcal{X}$ defined by a radius $r_{\text{cut}}$, a LLG, denoted as $\mathcal{G}_{\mathcal{X}} = (V,E)$, is composed by a set of vertices (nodes) $V = \left\lbrace v_{i} \right\rbrace_{i = 1}^{N}$ and a set of edges  $E = \left\lbrace e_{i} \right\rbrace_{i = 1}^{M}$ -- where $N$ and $M$
are associated with the number of atoms and the number of chemical bonds in the atomic environment, respectively \cite{ralaivola2005, kriege2020}. The LLG is said ``\textit{labeled}" if a label $l$ corresponding to the chemical symbol is attributed to each node $L(V)=\left\lbrace l(v_{i}) \right\rbrace_{i=1}^{N}$ (see Fig. \ref{Fig::LLG}). \\

A simple unlabeled graph is usually represented through an adjacency matrix $\mathbf{A} (\mathcal{G}_{\mathcal{X}}) \in \mathbb{R}^{N,N} $, 
whose elements $A_{ij}$ are set to one  if two different nodes $i$ and $j$ are linked by an edge or to zero otherwise \cite{nikolentzos2019, ralaivola2005}. However,  the same graph would correspond to an infinite number of conformations through this definition: by changing the interatomic distances, as long as the edges are preserved (\textit{i.e.} bonds are not broken), the corresponding graph will not change. This issue is particularly problematic for the prediction of total energy related properties  
(\textit{e.g.} atomization energy), which are strongly influenced by the precise atomic conformation.
For this reason in this work  we use weighted adjacency matrices~\cite{ferre2017} which include significantly more information on the structure of the atomic environment. Specifically, a Gaussian function is attributed to each atomic position $\mathbf{r}_{i}$ in the atomic environment. Instead of considering only binary values for the off-diagonal elements of the adjacency matrix, weights $w_{ij}$ between each pair of vertices $v_{i}$ and $v_{j}$ are set to correspond the overlap of these Gaussian functions, such that:
\begin{equation}
w_{ij}(r^{\text{cov}}_{i},r^{\text{cov}}_{j},\gamma)= \int \phi_{r^{\text{cov}}_{i}, \gamma} (\left\Vert \mathbf{r} - \mathbf{r}_{i} \right\Vert) \phi_{r^{\text{cov}}_{j}, \gamma} (\left\Vert \mathbf{r} - \mathbf{r}_{j} \right\Vert) \text{d} \mathbf{r} \; , 
\end{equation}
with $\phi_{r^{\text{cov}}_{i}, \gamma} (\left\Vert \mathbf{r} - \mathbf{r}_{i} \right\Vert) = \frac{1}{(2 \gamma \pi r^{\text{cov}}_{i})^{3/2}} \exp \left[ - \left( \frac{\left\Vert \mathbf{r} - \mathbf{r}_{i} \right\Vert }{ \sqrt{2 \gamma} {r^{\text{cov}}_{i}} } \right)^2 \right]$. 
Hence the weights depend on covalent radii $r^{\text{cov}}_i$ (coming from crystallographic data \cite{cordero2008}) of all elements and on the hyperparameter $\gamma$ controlling the width of the Gaussian atomic distributions. Finally, a threshold $\epsilon$ is applied as follows:
\begin{equation}\label{weighted_ad}
A_{ij} (\mathcal{G}_{\mathcal{X}}) = \left \lbrace 
\begin{array}{l l}
w_{ij} & \text{if} \; w_{ij} > \epsilon\\
0 & \text{otherwise}  
\end{array}
\right. \; .
\end{equation}
\textit{i.e.},  an edge  between the two nodes $v_{i}$ and $v_{j}$ is defined when $w_{ij}$ is larger than epsilon (see Fig. \ref{Fig::LLG}). These weights can be interpreted intuitively as probabilities to ``jump'' from one atom to another atom. 

\subsection{\label{sec::MPP} Maximum Probability Paths and Label Enrichment.}

From the definition of the weighted adjacency matrix $\mathbf{A}(\mathcal{G}_{\mathcal{X}})$ in Eq.~\ref{weighted_ad}, we can determine a path $\pi\left(v_{i},v_{j}\right)$ between each pair of vertices $\left(v_{i},v_{j}\right)$ (we suppose here that each node is connected to at least one another node, as it is the case for all the systems studied in this work). This path is defined by a finite-length sequence of vertices $\left(v_{i},\dots,v_{k},\dots,v_{j}\right)$ with the property that $\left(v_{k},v_{k+1}\right) \in E$ and $k = i, \dots, j-1$ \cite{mahe2004, borgwardt2005, mahe2005}. In this paper we will consider a machine learning approach based on a kernel that measures the similarity between paths. Specifically, we will focus on a variant of the shortest path (SP) kernel. While determining all paths of a graph is a NP-hard problem, computing the shortest path between pairs of vertices is a problem solvable in polynomial time ($\mathcal{O}(n^3)$) \cite{borgwardt2005}. \\ 

Typically, the determination of the shortest path is based on the distance and on the triangular inequality (Floyd-Warshall algorithm) \cite{floyd1962}. Rather then directly using an approach based on shortest paths, in the context of the present work we found more natural to introduce the analogous concept of maximum probability path (MPP). In particular, interpreting the elements of the weighted adjacency matrix as transition probabilities between two nodes, a path is identified with the greatest overall probability expressed as the product of the weights associated with all the edges that are involved in the path. Mathematically, the probability of a certain path can be written as:
\begin{equation}
\pi\left(v_{i},v_{j}\right) = \prod_{k=i}^{j-1} w_{k,k+1} \; ,
\end{equation}
where the intermediate indexes in the product correspond to all the nodes visited along the path.
In order to find the MPP between two vertices, we modified the original Floyd-Warshall algorithm\cite{borgwardt2005} (see Alg. \ref{Alg::FWT}).

\begin{algorithm}[!h]
\caption{\label{Alg::FWT}The Floyd-Warshall algorithm for MPPs}
	\begin{algorithmic}
		\REQUIRE Weighted adjacency matrix $\mathbf{A}(\mathcal{G}_{\mathcal{X}})$
		\ENSURE MPPs matrix $\mathbf{P}(\mathcal{G}_{\mathcal{X}})$
		
		\FOR {k = 1 to N} 
			\FOR {i = 1 to N}
				\FOR {j = 1 to N}
					\IF {A[i,k]*A[k,j] $>$ A[i,j]}
					
						\STATE A[i,j] = A[i,k]*A[k,j]
						
					\ENDIF
				\ENDFOR
			\ENDFOR
		\ENDFOR 
	\end{algorithmic}
\end{algorithm}

\begin{figure}[!h]
\includegraphics[scale=0.7]{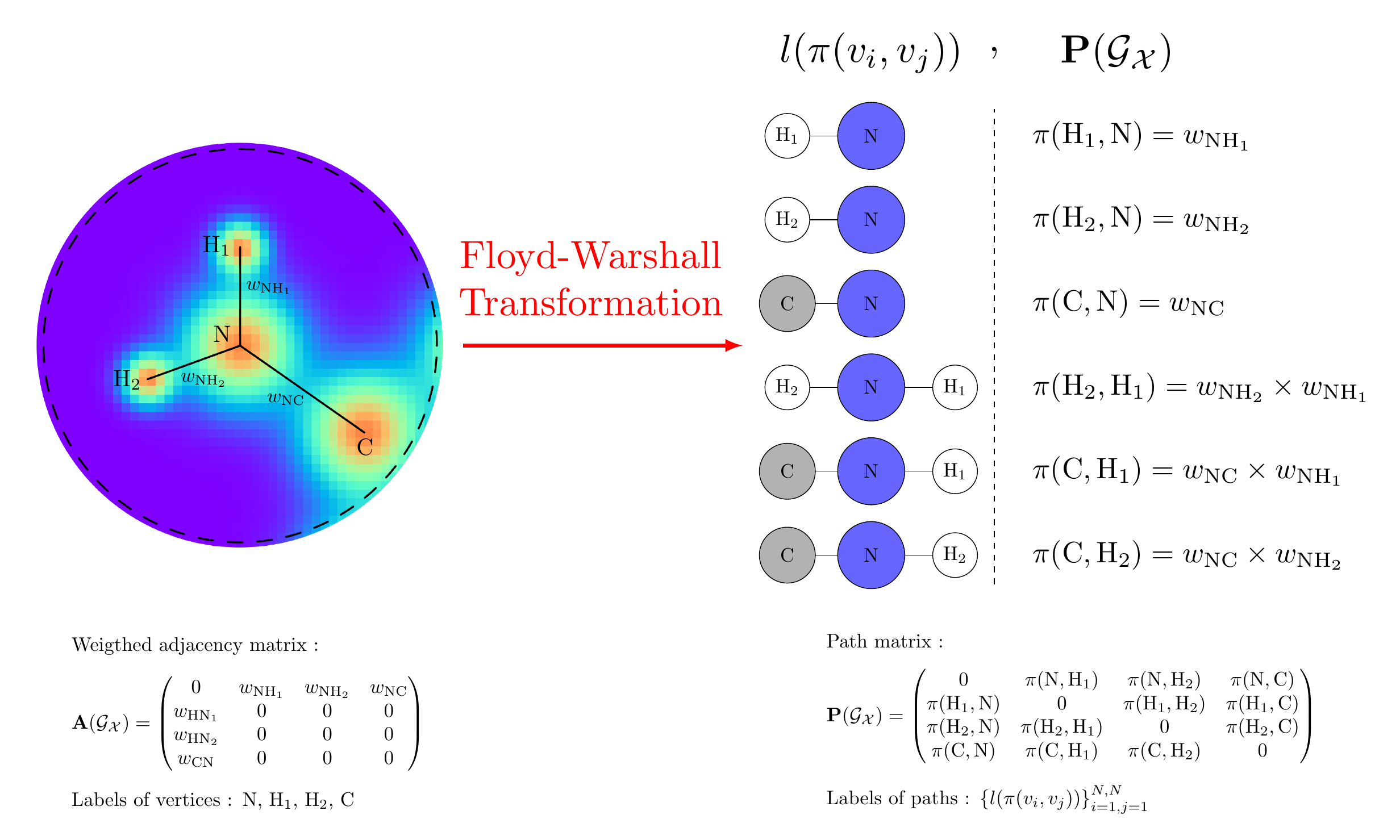}
\caption{\label{Fig::FW-Trans}Illustration of the Floyd-Warshall transformation. The weighted adjacency matrix $\mathbf{A}(\mathcal{G}_{\mathcal{X}})$ is transformed into a path matrix $\mathbf{P}(\mathcal{G}_{\mathcal{X}})$. Each path between pairs of atoms is characterized by a sequence of labels $l(\pi(v_{i},v_{j}))$ and a maximum probability path $\pi(v_{i},v_{j})$.}
\end{figure}

Through our variant of the Floyd-Warshall algorithm, the matrix $\mathbf{A}(\mathcal{G}_{\mathcal{X}})$ is transformed into a new matrix $\mathbf{P}(\mathcal{G}_{\mathcal{X}})$ that can be interpreted as a fully connected graph in which the matrix elements $P_{ij}$ correspond to maximized overall probabilities of transition from the node $i$ to $j$. In our method, the MPPs are also labeled by a sequence of labels $l(\pi(v_{i},v_{j})) = \left( l(v_{i}),\dots,l(v_{k}),\dots,l(v_{j})\right)$. As discussed below (see Sec. \ref{sec::HybMMPPKernel} A.III), 
this is necessary in order to include information on atomic species so that  only the MPPs  involving the same sequences of atoms are compared (see Fig. \ref{Fig::FW-Trans}).

\begin{figure}[!h]
\includegraphics[scale=1.5]{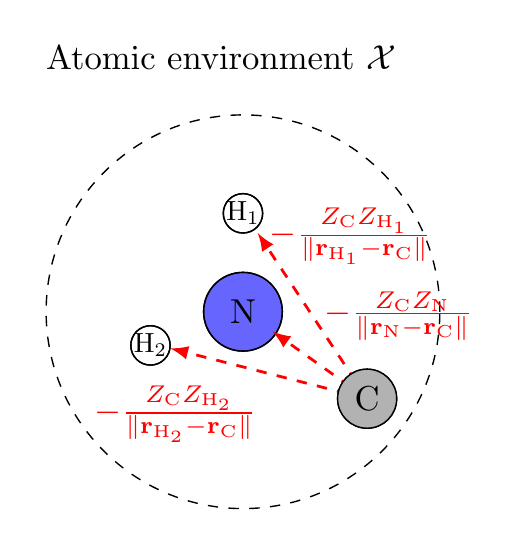}
\caption{\label{Fig::Enrichment}Illustration of the labels enrichment of nodes. In the atomic environment $\mathcal{X}$ to each node it is associated a sorted Coulomb vector $\mathbf{c}^{(i)}$. Example with the carbon atom : $\mathbf{c}^{(\text{C})} = \left\lbrace \frac{Z_{\text{C}} Z_{\text{N}}}{ \left\Vert \mathbf{r}_{\text{N}} - \mathbf{r}_{\text{C}} \right\Vert }, \frac{Z_{\text{C}} Z_{\text{H}_{1}}}{ \left\Vert \mathbf{r}_{\text{H}_{1}} - \mathbf{r}_{\text{C}} \right\Vert }, \frac{Z_{\text{C}} Z_{\text{H}_{2}}}{ \left\Vert \mathbf{r}_{\text{H}_{2}} - \mathbf{r}_{\text{C}} \right\Vert } \right\rbrace$. }
\end{figure}

The concepts described up to this point (weighted graphs and MPPs) allow for a rather accurate description of geometric motifs and bonding between atoms. However, these approaches are not sufficient to capture all the geometric details in atomic environments. Specifically, modifications (\textit{e.g.} rotations) which preserve interatomic bond distances in the local environment leave the matrices $\mathbf{A}(\mathcal{G}_{\mathcal{X}})$ and $\mathbf{P}(\mathcal{G}_{\mathcal{X}})$ unchanged. These finer structural details, which are not captured by the MPP approach, might have different relevance depending on the specific dataset but, nevertheless, contribute to the non-uniqueness of the connection between localized graph and property to predict.
To overcome this difficulty, we introduced a label enrichment of the vertices \cite{mahe2004, mahe2005}. 
Inspired by the Coulomb matrix descriptor introduced by Rupp \textit{et al.} \cite{rupp2012}, we defined for each node a Coulomb vector $\mathbf{c}^{(i)}$ sorted according to the distances in increasing order whose components are defined as
\begin{equation}
c_{j}^{(i)} = \frac{Z_{i} Z_{j}}{ \left\Vert \mathbf{r}_{j} - \mathbf{r}_{i} \right\Vert } \; ,
\end{equation}
where $Z_{i}$, $Z_{j}$ are the nuclear charges. The superscript $i$ refers to the node which is labeled, while the subscript $j$ is an index that indicates all the other atoms in the environment of $i$  (see Fig. \ref{Fig::Enrichment}). 

Because the atomic composition of environments can be different, to prevent any issue with dimensionality mismatch, the Coulomb vectors are padded with zeros to set their dimension to be equal and transferable. Finally, we obtain a new set of labels  $C_{\mathcal{X}} = \left\lbrace \mathbf{c}^{(i)} \right\rbrace_{i=1}^{N}$ that takes into account the missing geometric information in the description of atomic environments.

\subsection{\label{sec::HybMMPPKernel}The Hybrid Maximum Probability Path (HMPP) Kernel.}

The main idea of kernel approaches consists in finding an accurate measure of the similarity between pairs of data points (atomic environments in our case). For example, in the SOAP kernel, the similarity is determined through an overlap calculation of the atomic density of two atomic environments \cite{bartok2013, de2016}. When normalized, the kernel is close to one when two atomic environments are nearly identical and tends to zero if they are dissimilar. \\

\begin{figure}
\includegraphics[scale=0.7]{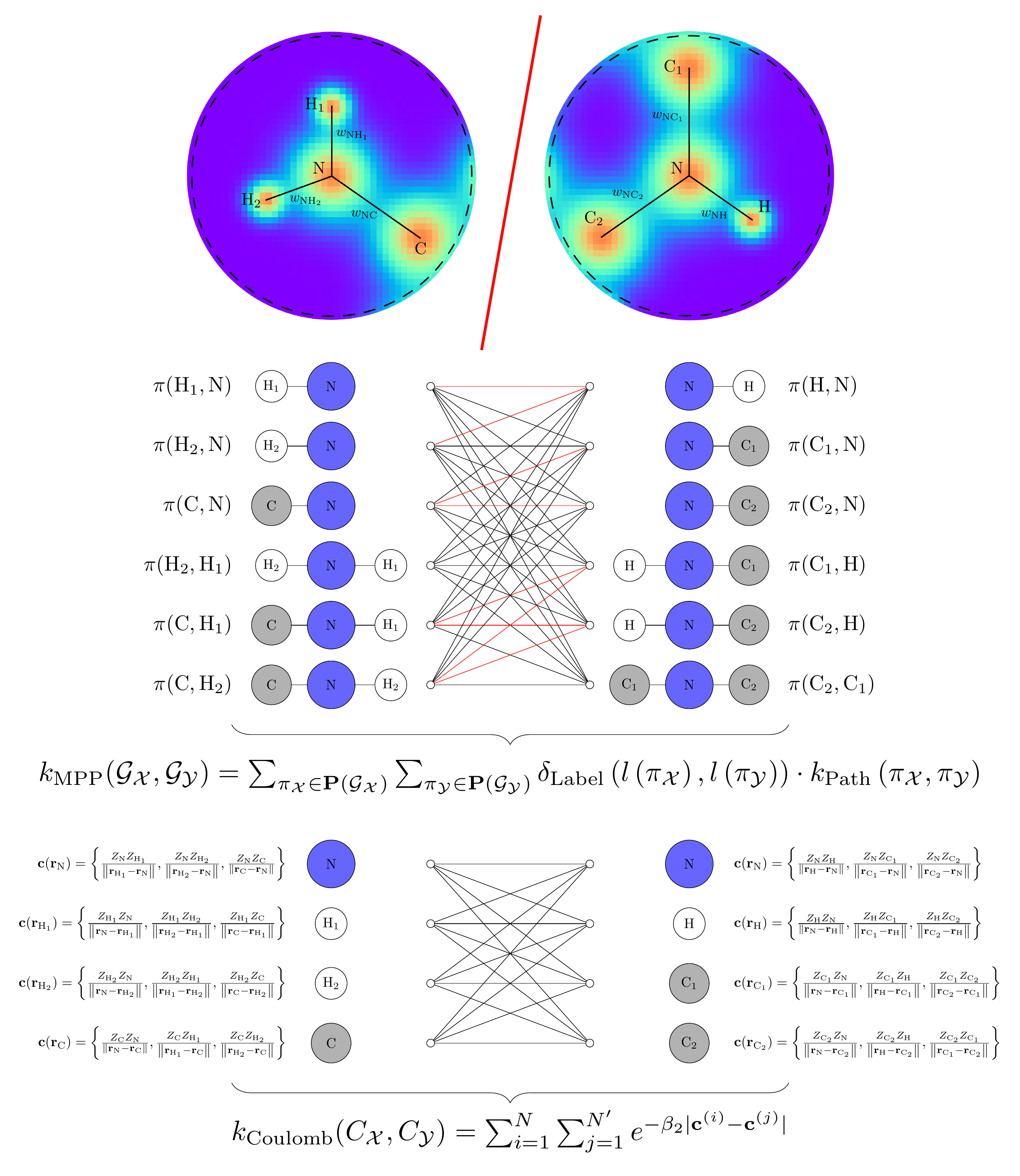}
\caption{\label{Fig::HMMPP}Illustration of the hybrid maximum probability path kernel. The calculation is based on two similarity measurements, one on the paths of atomic environments and the other one on the Coulomb labels of the nodes.}
\end{figure}

In our approach, an atomic environment $\mathcal{X}$ is characterized by a set of MPPs between all pairs of nodes $P_{ij}(\mathcal{G}_{\mathcal{X}}) = \pi \left( v_{i}, v_{j} \right)$, 
with their respective labels $L(\mathbf{P}(\mathcal{G}_{\mathcal{X}})) = \left\lbrace l(\pi(v_{i},v_{j})) \right\rbrace_{i=1,j=1}^{N,N}$ and a set of Coulomb vectors $C_{\mathcal{X}} = \left\lbrace \mathbf{c}^{(i)} \right\rbrace_{i=1}^{N}$
(see Sec~\ref{sec::MPP}). 
Being inspired by the iterative similarity for molecular graphs introduced by Rupp \textit{et al.} \cite{rupp2007}, where the use of a linear combination of two  kernels applied to vertices and edges was proposed,
the following hybrid kernel is introduced to compare the atomic environments (see also Fig.~\ref{Fig::HMMPP}):
\begin{equation}\label{eq:hybrid}
k(\mathcal{X},\mathcal{Y}) = (1-\alpha)k_{\text{MPP}}(\mathcal{G}_{\mathcal{X}}, \mathcal{G}_{\mathcal{Y}}) + \alpha k_{\text{Coulomb}} (C_{\mathcal{X}}, C_{\mathcal{Y}}) \; .
\end{equation}
This kernel is normalized via Tanimoto's normalization \cite{ralaivola2005}: 
\begin{equation}
k(\mathcal{X},\mathcal{Y}) = \frac{k(\mathcal{X},\mathcal{Y})}{k(\mathcal{X},\mathcal{X}) + k(\mathcal{Y},\mathcal{Y}) - k(\mathcal{X},\mathcal{Y})} \; ,
\end{equation}
which is commonly used also in other graphs application \cite{ralaivola2005, nikolentzos2017, nikolentzos2019, kriege2020}. 
This normalization can be seen as a Jaccard's distance -- \textit{i.e.} measuring how two sets intersect -- between two discrete collections of paths and vertices.  
The kernel $k(\mathcal{X},\mathcal{Y})$ is positive semidefinite \cite{nikolentzos2017}, as required for use in kernel-based machine
learning algorithms. 

To compare graph paths, we have chosen an approach similar to the labeled shortest path graph kernel \citep{borgwardt2005}. 
In its original definition this kernel could take values different from 0 only when applied to pairs of (shortest) paths with the same initial and final labels \citep{borgwardt2005}. In this work we found that this formulation does not reach a satisfactory level of accuracy. Indeed, even if the initial and final nodes of a pair of MPPs have the same ``atomic'' labels (namely they correspond to the same atomic species),  
the intermediate paths could be significantly different from a chemical point of view (\textit{i.e.} involve different elements) and still be considered as similar according to this original definition of the kernel. For this reason in the present work we used an MPP kernel which is strictly 0 unless the two paths to be compared have exactly   
the same sequence of atomic labels $l(\pi)$. This kernel can be expressed as follows: 
\begin{equation}
k_{\text{MPP}} (\mathcal{G}_{\mathcal{X}}, \mathcal{G}_{\mathcal{Y}}) = \sum_{\pi_{\mathcal{X}} \in \mathbf{P}(\mathcal{G}_{\mathcal{X}})} \sum_{\pi_{\mathcal{Y}} \in \mathbf{P}(\mathcal{G}_{\mathcal{Y}})} \delta_{\text{Label}} \left( l \left( \pi_{\mathcal{X}} \right), l \left( \pi_{\mathcal{Y}} \right) \right) \cdot k_{\text{Path}} \left( \pi_{\mathcal{X}}, \pi_{\mathcal{Y}}
 \right) \; ,
\end{equation}
where the $\delta_{\text{Label}} \left( l \left( \pi_{\mathcal{X}} \right), l \left( \pi_{\mathcal{Y}} \right) \right)$ is a 
delta kernel applied on the path labels such that:
\begin{equation}
\delta_{\text{Label}} \left( l \left( \pi_{\mathcal{X}} \right), l \left( \pi_{\mathcal{Y}} \right) \right) = \left\lbrace 
\begin{array}{l l}
1 & \text{if} \; l \left( \pi_{\mathcal{X}} \right) = l \left( \pi_{\mathcal{Y}} \right) \\
0 & \text{otherwise} \\
\end{array}
\right. \; .
\end{equation}
This kernel is equal to one if, and only if, the sequences of labels along $\pi_{\mathcal{X}}$ and $\pi_{\mathcal{Y}}$ are identical. The term  $k_{\text{Path}} \left( \pi_{\mathcal{X}}, \pi_{\mathcal{Y}} \right)$ is the Laplacian kernel \cite{rupp2015} 
\begin{equation}
k_{\text{Path}} \left( \pi_{\mathcal{X}}, \pi_{\mathcal{Y}} \right) = e^{-\beta_{1} \left\vert \pi_{\mathcal{X}} - \pi_{\mathcal{Y}} \right\vert} \; ,
\end{equation} 
used to compare pairs of MPPs. Similarly, the Coulomb labels are compared through a sum of Laplacian kernel functions: 
\begin{equation}
k_{\text{Coulomb}} (C_{\mathcal{X}}, C_{\mathcal{Y}}) = \sum_{i=1}^{N} \sum_{j=1}^{N'} e^{-\beta_{2} \left\vert  \mathbf{c}^{(i)} - \mathbf{c}^{(j)} \right\vert} \; , 
\end{equation} 
where $N$ and $N'$ denote the number of atoms in the atomic environments $\mathcal{X}$ and $\mathcal{Y}$, respectively, and $\beta_1$ and $\beta_2$ 
are the hyperparameters that control the decay of the Laplacian kernel functions. They play a central role in the prediction quality and are estimated through a grid search using a validation dataset. 
The relative importance of the two kernels $k_{\text{MPP}}$ and $k_{\text{Coulomb}}$ in the description of a given system 
is controlled by the hyperparameter $\alpha$ in Eq.~\ref{eq:hybrid}. 


\section{\label{sec::Benchmarks}Benchmarks}

To demonstrate the accuracy of our hybrid graph kernel approach we applied it to two regression problems of energy-related properties and compared its performance to that of the popular SOAP descriptor \cite{bartok2013,de2016} and graph approximated energy (GRAPE) \cite{ferre2017}. The machine learning models were based on  in-house programs written in {\tt{Python3}} employing the {\tt{DScribe}} \cite{dscribe} and the {\tt{SciKit-Learn}} \cite{scikit-learn} libraries implementing the SOAP descriptors and KRR routines, respectively.     

\subsection{The QM7 dataset.}\label{ssec::qm7}

The QM7 dataset consists of atomization energies for a total of 7165 organic molecules \cite{rappe1992, blum2009, rupp2012}. The individual molecules are composed of up to 23 atoms and contain at most five different elements involving H, C, N, O and S. The atomization energies, which span a range between -2000 to -800 kcal/mol, have been computed using density functional theory (DFT) \cite{kohn1964} within the Perdew-Burke-Ernzerhof (PBE) \cite{pbe} parameterization of the generalized gradient approximation. \\ 

Using the five-fold partitioning originally proposed by Rupp \textit{et al.} \cite{rupp2012} and  following the work of Ferr\'e \textit{et al.} \cite{ferre2017}, we considered the first $N$ molecules (with $N=100$, $300$, $500$ and $1000$ points) of the first fold as a training set. The second and third folds were combined to create a validation set for the grid search of the optimal hyperparameters, which, in the case of the HMPP kernel
include the parameters
$r_{\text{cut}}$, $\gamma$, $\beta_{1}$, $\beta_{2}$ and $\alpha$. 
Finally, the two last folds were used as a test set to evaluate the accuracy of the model through the mean absolute error (MAE) and the root mean square error (RMSE) of predicted energies. 

\subsection{The BA10 dataset.} 

The BA10 dataset contains standard enthalpies of formation for a set of ten binary alloys
(AgCu, AlFe, AlMg, AlNi, AlTi, CoNi, CuFe, CuNi, FeV and NbNi) represented by 1595 configurations each \cite{nyshadam2019} (\textit{i.e.} altogether 15950 configurations). 
The corresponding alloys are obtained considering all the possible unit cells with 1 to 8 atoms 
for the face-centered cubic (fcc) and body-centered cubic (bcc) Bravais lattices, and all the possible cells with 2 to 8 atoms for the hexagonal close-packed (hcp) symmetry. The crystal structures were determined using the Hart and Forcade algorithm~\cite{hart:2012} and the lattice parameters were set according to the Vergard's law \cite{nyshadam2019}. Finally, the standard enthalpies of formation were computed through the Vienna Ab initio Simulation Package \cite{vasp_1, vasp_2} (VASP) at the DFT/PBE \cite{pbe} 
level of theory without geometric relaxation. More details on this dataset can be found in the original article of Nyshadham \textit{et al.} \cite{nyshadam2019}. \\
 

The hyperparameters were optimized on the configurations of the alloy AgCu. Specifically, by training the machine learning models on 100 randomly chosen points  we determined the optimal hyperparameters on an independent validation set composed by other 100 randomly chosen structures. The set of hyperparameters determined for AgCu has then been used for all the other compounds. This procedure allowed us to evaluate the transferability of the hyperparameters into the alloy space. Within this approach we have studied for each alloy the influence of the training set size on the quality of predictions. To this end, we considered four different training sets with $N = 100$, 300, 500 and 1000 configurations. As in the case of the QM7 dataset (see Sec.~\ref{ssec::qm7}), the regression quality was quantified trough the MAE and the RMSE
evaluated for the predicted energies on a test set composed by the rest of the structures (\textit{i.e.} $1595 - N$ configurations per alloy). 
 
\section{\label{sec::Results} Results and Discussion} 

\subsection{Optimization of the hyperparameters}

The determination of the hyperparameters was conducted on a hold-out validation set through a five-dimensional grid search for the parameters $r_{\text{cut}}$, $\gamma$, $\beta_{1}$, $\beta_{2}$, and $\alpha$. The cutoff radius $r_{\text{cut}}$, that defines the size of each atomic environment, was evaluated in the interval between 1.6 and 2.4 \AA{} with an increment of 0.2 \AA{} for the QM7 dataset, 
and between 2.0 to 5.0 \AA{} with an increment of 1.0 \AA{} for the BA10 dataset. 
The values of 1.8 \AA{} and 3.0 \AA{} were identified as optimal for the QM7 and BA10 datasets, respectively. 
The scaling factor $\gamma$, that governs the width of the atomic
Gaussian functions, was tested in the range from 0.1 to 0.9 \AA{} with a step of 0.2 \AA{}; the optimal values of 0.5 \AA{} and 0.5 \AA{} have been 
identified for the QM7 and BA10 datasets, respectively. 

The two hyperparameters $\beta_{1}$ and $\beta_{2}$ that define the decay rate of the two exponential functions in the $k_{\text{MPP}}$ and $k_{\text{Coulomb}}$ kernels were evaluated on a decimal logarithmic grid defined on the interval between $10^{-1}$ to $10^{-4}$. For both the datasets considered here, the optimal hyperparameters values were found to be 0.1 and 0.001 for $\beta_1$ and $\beta_2$, respectively. Compared to the other three parameters in the model, $\beta_{1}$ and $\beta_{2}$ seem to have a weaker system dependence and could be possibly transferred across different datasets. Although this observation should be confirmed by future investigations on several different datasets, the possibility of fixing $\beta_{1}$ and $\beta_{2}$ a priori could help to simplify the hyperparameter optimization procedure.

The $\alpha$ parameter, that describes the relative contribution of the kernels $k_{\text{MPP}}$ and $k_{\text{Coulomb}}$, was tuned by considering values between 0 to 1 with a step of 0.1. For the QM7 dataset, a relatively small $\alpha$ (0.2) exhibited the best performance, while a significantly higher value 
(0.9) was identified for the dataset BA10. It is important to notice that the hybridization brings important improvements with respect to the sole use of the best performing between the $k_{\text{MPP}}$ and $k_\text{Coulomb}$ kernels (for the smallest training sets considered below the hybridization lowers the mean absolute error by $35$ \% and $17$ \% for QM7 and BA10, respectively).  


\subsection{Prediction of the atomization energies.}

\begin{figure}[!h]
\begin{tabular}{c c}
\includegraphics[scale=0.6]{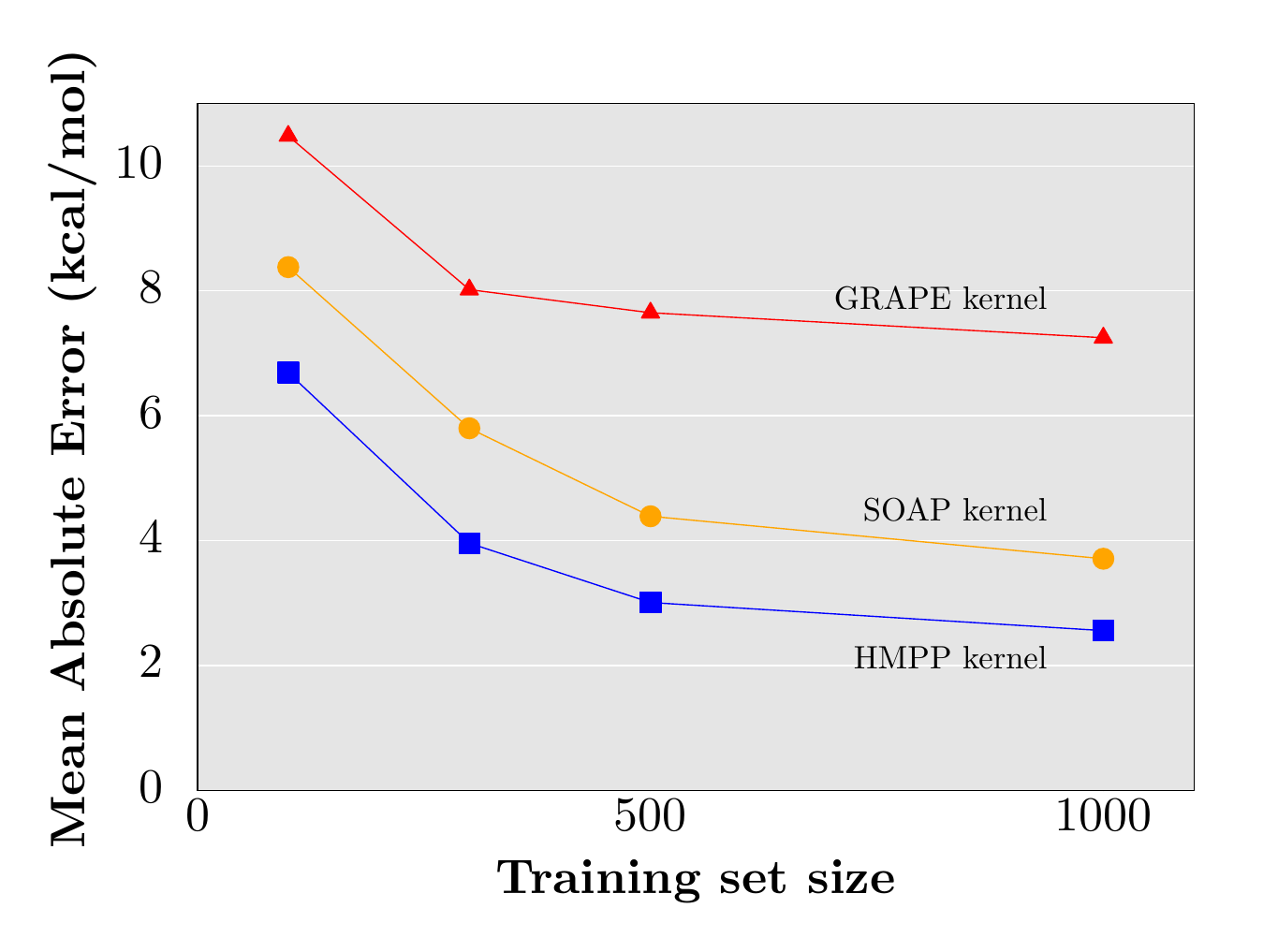} & 
\includegraphics[scale=0.6]{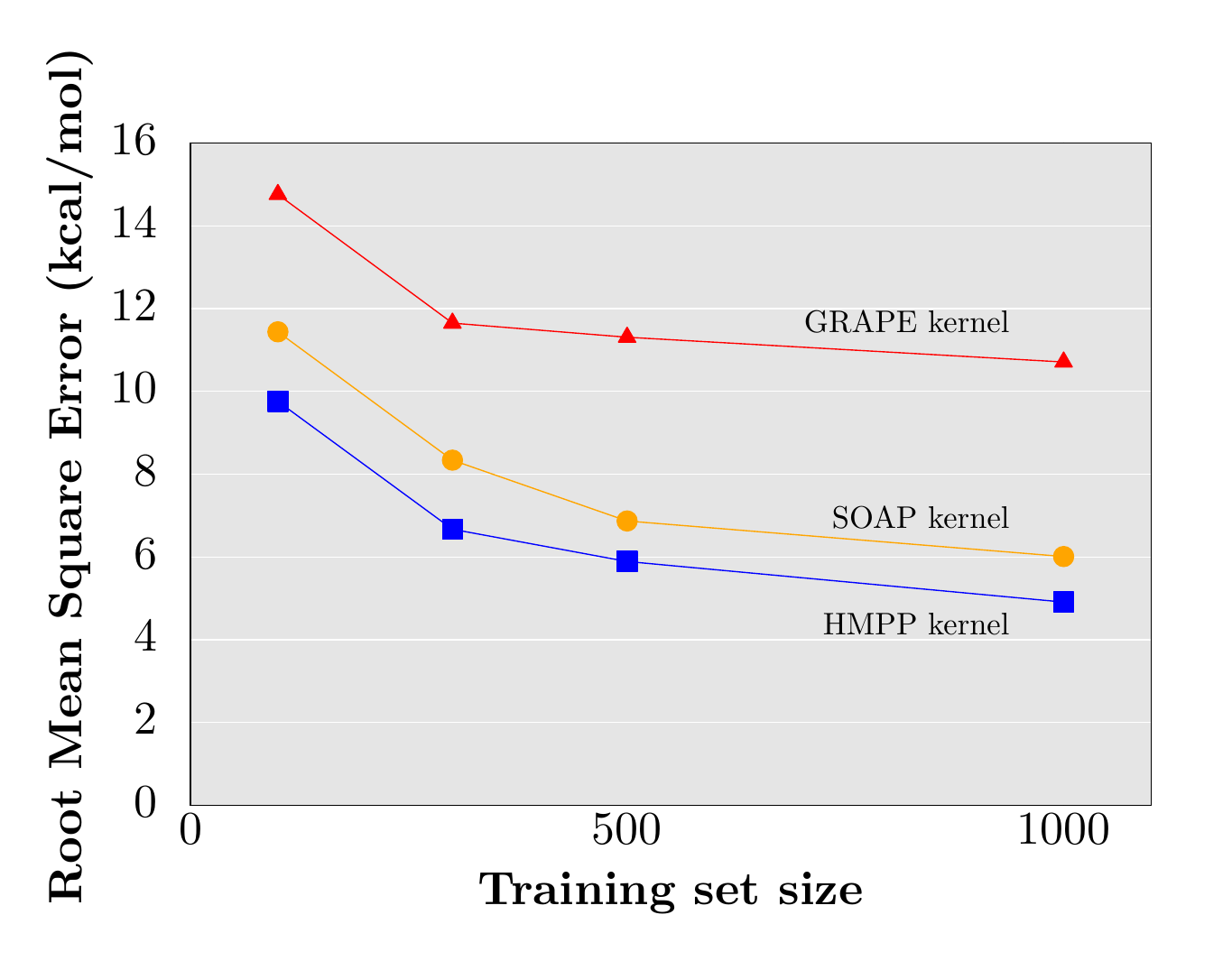} \\
(a) MAE & (b) RMSE
\end{tabular}
\caption{\label{Fig::QM7}Variation of the quality of predictions of atomization energies (QM7 dataset) of the three kernels studied in this work (SOAP, GRAPE and HMPP) with the training set size as measured by the mean absolute error (MAE) and the root mean square error (RMSE). The values are given in kcal/mol (See also Table \ref{Tab::QM7-TrSet}).}
\end{figure}

In this section we test the performance of the HMPP kernel in predicting the atomization energies of the molecules in the QM7 dataset. Table \ref{Tab::QM7-TrSet} and Figure \ref{Fig::QM7} show the variation of the MAEs and RMSEs as a function of the training set size. For the sake of comparison we also show results for some other well established approaches, namely the SOAP kernel and  GRAPE, the latter being a localized graph kernel method previously reported in the literature \cite{ferre2017} and applied to  regression problems. The results for the different methods have been generated using the same training sets and analogous procedures for the optimization of the hyperparameters. We observe that the HMPP kernel outperforms the other two methods regardless of the training set size. It is important to notice that in Ref. 37 the results reported for the GRAPE kernel  slightly improved over the SOAP method (which is in contrast to our results from Table \ref{Tab::QM7-TrSet} and Figure \ref{Fig::QM7}). However, the authors of this work clearly stated that the tuning of the hyperparameters was limited and that, accordingly, it was not possible to conclude ``that one method outperforms the other''. The MAEs and RMSEs reported in Table \ref{Tab::QM7-TrSet} for GRAPE and SOAP are also sizeably smaller than those in Ref. 37, demonstrating the importance of a fine tuning of the model hyperparameters to fully establish the accuracy of a certain approach.

\begin{table}[!h]
\begin{tabular}{l c c c c}
\hline
	                 & \textbf{100 points} & \textbf{300 points} & \textbf{500 points} & \textbf{1000 points} \\
\hline
\hline
	\textbf{SOAP}    &            &            &            &             \\
	MAE              & 8.38       &	5.80       & 4.39       & 3.71        \\
	RMSE             & 11.44	  & 8.34	   & 6.87	    & 6.01        \\
\hline
	\textbf{GRAPE}   &            &            &            &             \\
	MAE              & 10.48      &	8.02       & 7.65       & 7.25        \\
	RMSE             & 14.76      &	11.65      & 11.31      & 10.71       \\
\hline
	\textbf{HMPP}   &              &            &            &             \\
	MAE               & 6.69         &	3.96       & 3.01       & 2.56        \\
	RMSE              & 9.76	      & 6.67       & 5.89       & 4.91        \\
\hline
\end{tabular}
\caption{Variation of mean absolute error (MAE) and the root mean square error (RMSE) (in kcal/mol) of atomization energies (QM7 dataset) obtained using the SOAP, GRAPE and HMPP kernels with the training set size (cf. Figure \ref{Fig::QM7}).}
\label{Tab::QM7-TrSet}
\end{table}

The GRAPE kernel uses a random walk kernel directly applied on the weighted adjacency matrix. 
Within this approach, each component of the adjacency matrix is defined through an overlap of atomic Gaussian functions of the same width. However, we can observe that this use of graphs 
does not reach a satisfactory level of accuracy, as shown by the high MAE and RMSE values in Table \ref{Tab::QM7-TrSet} and Figure \ref{Fig::QM7}. This is likely related to the lack of labels in the GRAPE approach. Tang and de Jong 
have proposed in 2019 a marginalized graph kernel to predict the atomization energies and  have obtained impressive results with the introduction of labels \cite{tang2019}. However, the reported model was based on a global description of the molecular graphs and, differently from our present approach, the corresponding graph kernel cannot be applied  to solid state systems in a straightforward way. \\

In Table \ref{Tab::QM7-Large} we show the values of MAE and RMSE obtained for larger training set sizes. Specifically, two additional training sets with 2000 and 5000 data points were built through a random sampling without stratification of the data. In this table, the results that we obtained for the HMPP, SOAP, and GRAPE kernel methods are compared with some other results presented in the literature obtained using different kernels and/or descriptors. Since different training sets are used in different papers, such a comparison can be considered as only qualitative. 
Nevertheless, the results compiled in Table \ref{Tab::QM7-Large} clearly show  that the HMPP approach is competitive with more traditional approaches. 
In particular, our HMPP kernel is more accurate than the GRAPE, SOAP (average kernel AK), and Coulomb matrix (CM) approaches while its performance is comparable  to the global molecular graph (GMG) \cite{tang2019} and 
bag of bonds (BoB) \cite{hansen2015} methods. It is important to further stress that comparison of our results with those produced by different groups can be only qualitative, as  a stratification of the data was used for most of the results in the literature and, in the case of the BoB, an even larger training set was considered (5732 data points). \\

\begin{table}[!h]
\begin{tabular}{l l l l c c l}
\hline 
Training set & Representation & Kernel & Regression & MAE & RMSE & Source \\ 
\hline 
\hline
\textbf{2000, random}     & \textbf{HMPP}     & \textbf{LGK}                    & \textbf{KRR} & \textbf{1.87} & \textbf{4.17}     & \textbf{This work}              \\
2000, random     & GRAPE           & LGK                    & KRR & 7.00 & 10.21    & This work              \\
2000, random     & SOAP            & AK                     & KRR & 2.95 & 4.61     & This work              \\
\hdashline
2000, random     & GMG             & GK              & GPR & 1.48 & 3.57     & \cite{tang2019}  \\
2000, stratified & CM              & Laplacian              & KRR & 4.32 & $\cdots$ & \cite{hansen2013} \\
\hline
\textbf{5000, random}     & \textbf{HMPP}     & \textbf{LGK}                    & \textbf{KRR} & \textbf{1.59} & \textbf{3.07}     & \textbf{This work}              \\
5000, random     & GRAPE           & LGK                    & KRR & 6.99 & 10.05    & This work              \\
5000, random     & SOAP            & AK                     & KRR & 2.59 & 3.63     & This work              \\
\hdashline
5000, random     & GMG             & GK                     & GPR & 1.01 & 2.29     & \cite{tang2019}   \\
5000, stratified & SOAP            & REMatch                & KRR & 0.92 & 1.61     & \cite{de2016}     \\
5732, stratified & CM              & Laplacian              & KRR & 3.07 & 4.84     & \cite{hansen2013} \\  
5732, stratified & BoB             & Laplacian              & KRR & 1.50 & $\cdots$ & \cite{hansen2015} \\
\hline
\hline
\end{tabular}
\caption{MAEs and RMSEs (in kcal/mol) determined for atomization energy predictions (QM7 dataset) made using different  state-of-the-art methods trained on  large training sets. The results 
obtained using the HMPP kernel proposed in this work are in bold. \\ 
\textit{List of abbreviations: Hybrid maximum probability path (HMPP), Graph approximated energy (GRAPE), Smooth overlap of atomic positions (SOAP), Global molecular graph (GMG), Coulomb matrix (CM), Bag of Bond (BoB), (Localized) graph kernel ((L)GK), Average kernel (AK), Regularized entropy match (REMatch), Kernel ridge regression (KRR), Gaussian process regression (GPR)}.}
\label{Tab::QM7-Large}
\end{table}  

\subsection{Prediction of the standard enthalpies of formation.}

\begin{table}[!h]
\begin{tabular}{l c c c c}
\hline
	                 & \textbf{100 points} & \textbf{300 points} & \textbf{500 points} & \textbf{1000 points} \\
\hline
\hline
	\textbf{SOAP}      &              &            &            &             \\
	MAE                & 0.21         & 0.17	   & 0.16	    & 0.15        \\
	RMSE               & 0.28         & 0.22       & 0.21       & 0.20        \\
\hline
	\textbf{HMPP}      &              &            &            &             \\
	MAE                & 0.28         &	0.19       & 0.16       & 0.12        \\
	RMSE               & 0.39	      & 0.26       & 0.21       & 0.17  \\
\hline
\end{tabular}
\caption{Variation of MAEs and RMSEs (in kcal/mol) with the training set size 
for enthalpies of formation of the BA10 dataset predicted using the SOAP and HMPP kernels (cf. Figure \ref{Fig::BA10}).}
\label{Tab::BA10-TrSet}
\end{table}

\begin{figure}[!h]
\begin{tabular}{c c}
\includegraphics[scale=0.6]{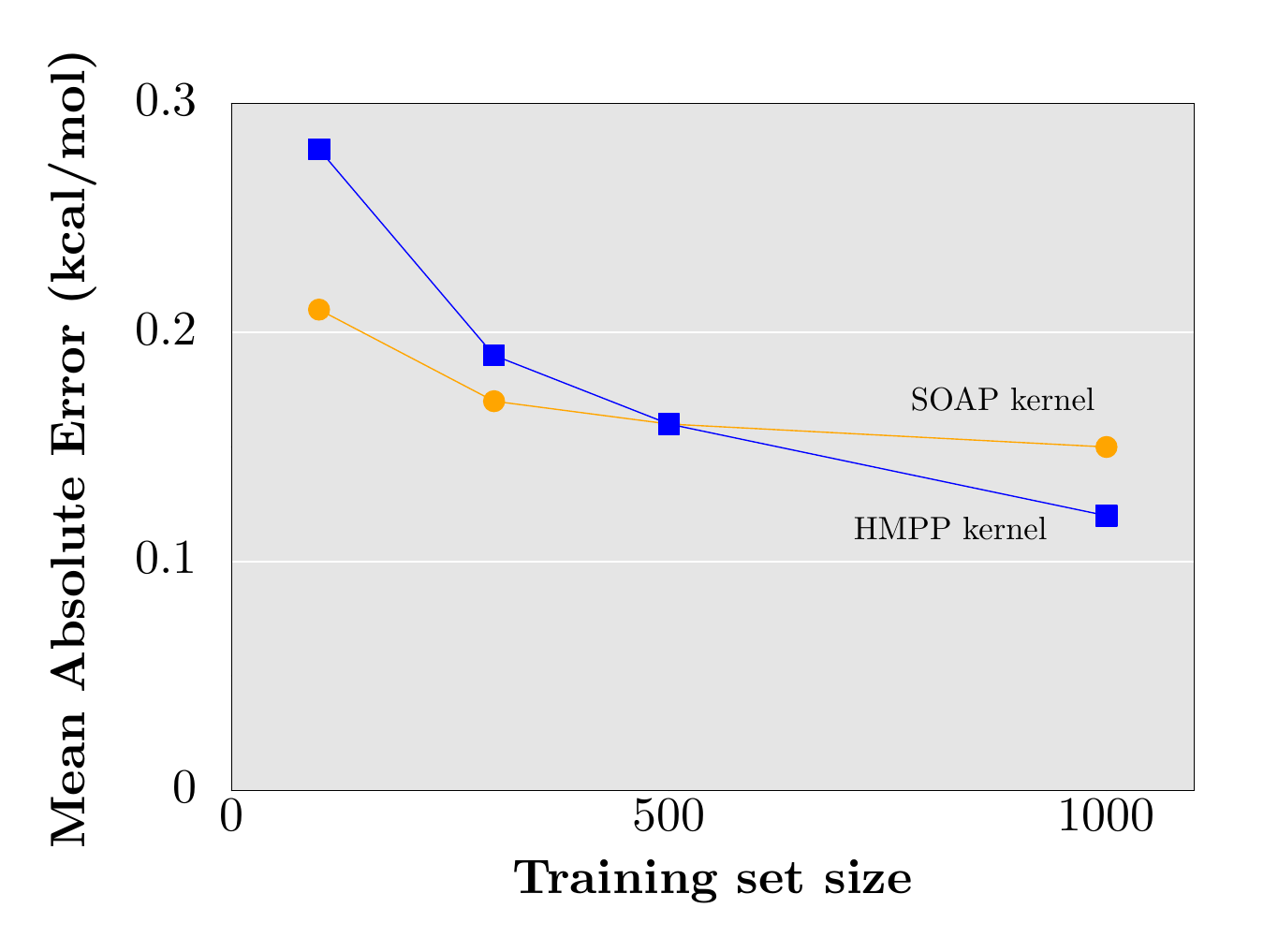} & 
\includegraphics[scale=0.6]{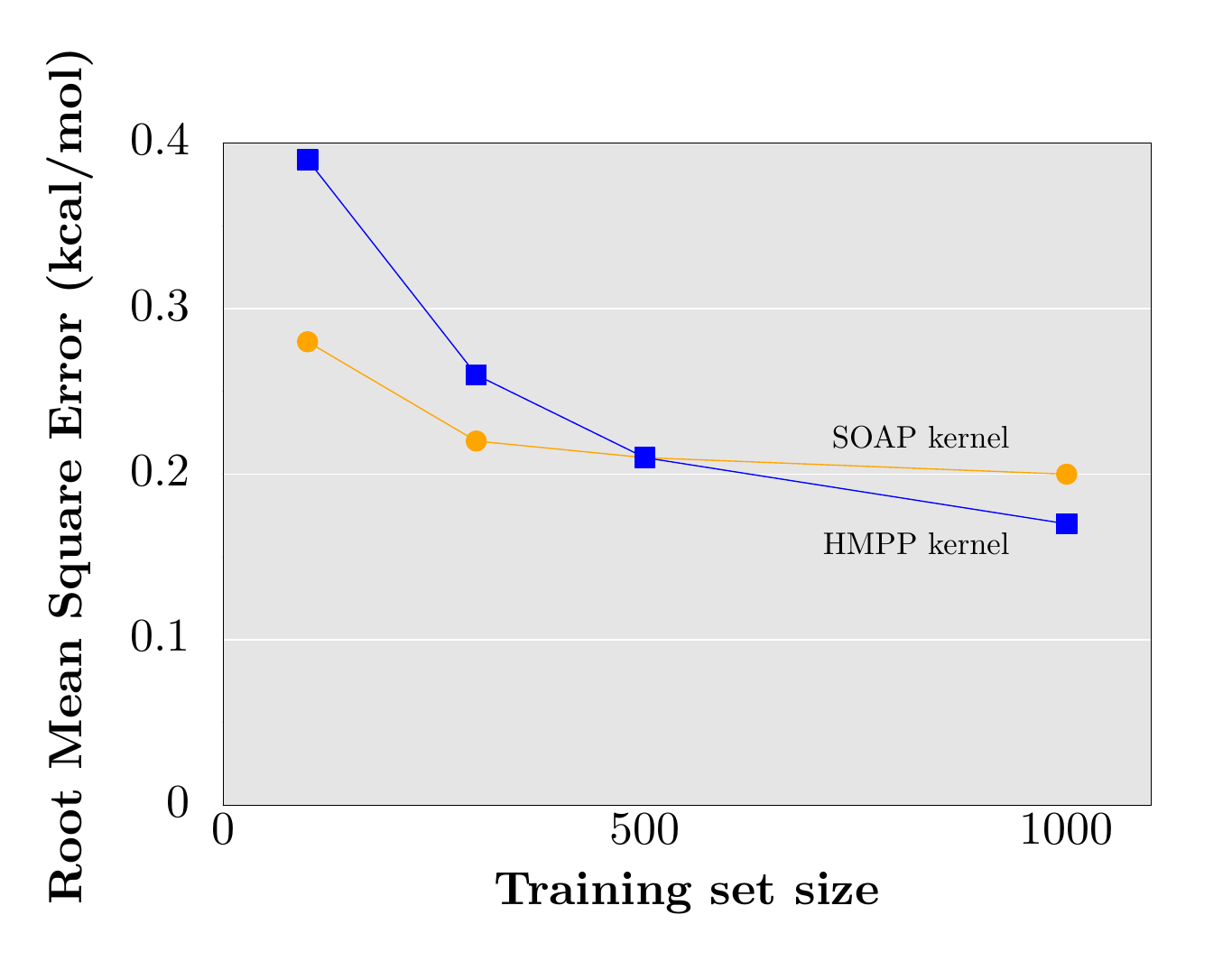} \\
(a) MAE & (b) RMSE
\end{tabular}
\caption{\label{Fig::BA10}Variation with the training set size of the MAEs and RMSEs for the formation enthalpy predictions (BA10 dataset) obtained from the SOAP and HMPP kernels. The values are given in kcal/mol (cf. Table \ref{Tab::BA10-TrSet}).}
\end{figure}

To evaluate the accuracy of the HMPP kernel model in the prediction of properties 
 of solid state systems, the entalpies of formation of the compounds in the BA10 dataset have been considered. Table \ref{Tab::BA10-TrSet} and Figure \ref{Fig::BA10} show the values of MAE and RMSE averaged over all the structures of the 10 binary alloys. The training sets, which are always excluded in the error evaluation, are obtained by randomly selecting 100, 300, 500, and 1000 configurations for each binary compound among the 1595 structures -- \textit{i.e.} without stratification of the data. As for the dataset QM7 discussed in the previous section (see Sec. \ref{sec::Results}), the present kernel was compared to the two other methods: SOAP and GRAPE. The SOAP descriptor has been used in the framework of the average kernel (AK), while the GRAPE has been employed in its original definition -- \textit{i.e.} in combination with a random walk graph kernel \cite{ferre2017}. For the sake of consistency,  hyperparameters used in all the methods have been obtained via the procedure explained in Sec. \ref{sec::Benchmarks} and the same training sets have been used in all calculations. The predictions of the localized graph kernel GRAPE were significantly worse than those obtained with SOAP and HMPP. For instance, with a training set of 100 points, the RMSE of the predicted enthalpies (averaged overall the alloys) is five times higher compared to the other two approaches. 
 Therefore, we omit the detailed discussion of results obtained with the GRAPE approach for the BA10 dataset.


It can be observed that the SOAP kernel performs slightly better than the hybrid MPP
 for the training sets of 100 and 300 configurations. However, the learning capability of the SOAP kernel seems to saturate already at 300 configurations, as apparent from the fact that the MAE and RMSE values do not decrease significantly by increasing the number of training data points beyond this value.
This could become a significant limitation if, for example, it would be of interest to extend the prediction of enthalpies on additional crystal structures. Our proposed model presents a better learning curve with errors that steadily decrease by increasing the training set size and become lower than SOAP for 1000 training structures. \\

\begin{figure}[!h]
\centering
\includegraphics[scale=0.6]{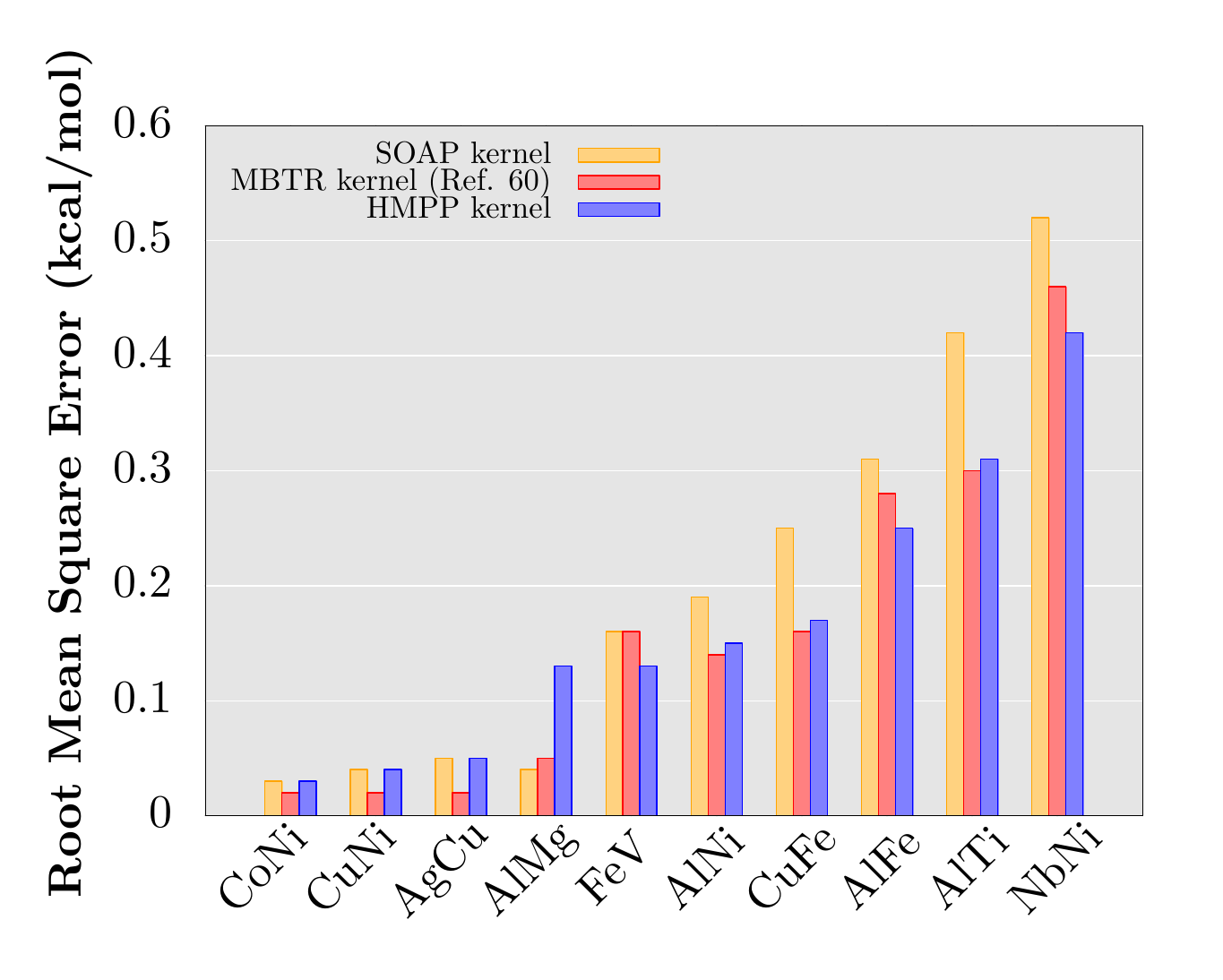} 
\caption{\label{Fig::BA10-Alloy}RMSE (in kcal/mol) of the SOAP, MBTR, and HMPP predictions of the standard enthalpies of formation for the ten alloys from the BA10 datased.}
\end{figure}

In Figure \ref{Fig::BA10-Alloy} we show the RMSE obtained for each of the ten binary alloys separately using predictions made by ML trained on 1000 points. It can be noticed that in most cases the RMSE obtained with the HMPP kernel is smaller compared to the SOAP approach. The only exception is represented by the AlMg alloy, as the RMSE of the HMPP kernel method is sizeably larger than that of SOAP  (in this comparison it should also be kept into account that hyperparameters for the HMPP model are not reoptimized for each alloy). In Figure \ref{Fig::BA10-Alloy} we also report results from the previous work of Nyshadham \textit{et al.} \cite{nyshadam2019} based on MBTR. For four binary systems, CoNi, CuNi, AgCu, and AlMg, the MBTR yields a particularly small RMSE. With the exception of AlMg,  the hybrid MPP approach provides a level of accuracy that is comparable to the previous MBTR calculations given in Ref. 60 and it even outperforms it in certain cases.


From Table \ref{Tab::BA10-resume}, we can conclude that our method reaches a high level of accuracy. Indeed, the HMPP kernel presents small values of MAE and RMSE and performs similarly to the MBTR kernel. Moreover, the difference with previous results reported in the literature based on the SOAP descriptor and the Gaussian process regression (GPR) is quite small (0.02 kcal/mol). Hence, like in the case of the QM7 dataset, we can conclude that our model describes properly also the BA10 dataset.

\begin{table}[!h]
\begin{tabular}{l l l l c c l}
\hline 
Training set & Representation & Kernel & Regression & MAE & RMSE & Source \\
\hline 
\hline
\textbf{1000}    & \textbf{HMPP}     & \textbf{LGK}  & \textbf{KRR} & \textbf{0.12}    & \textbf{0.17}      & \textbf{This work}              \\
1000             & SOAP            & AK                     & KRR & 0.15    & 0.20    & This work              \\
\hdashline
1000             & MBTR            & Laplacian              & KRR & 0.12    & $\cdots$ & \cite{nyshadam2019}   \\
1000             & SOAP            & $\cdots$               & GPR & 0.10    & $\cdots$ & \cite{nyshadam2019}   \\
\hline
\end{tabular}
\caption{MAE and RMSE for the HMPP, SOAP and MBTR predictions averaged over all  ten binary alloys from the BA10 dataset. The values are given in kcal/mol. \\
\textit{List of abbreviations: Hybrid maximum probability path (HMPP), Smooth overlap of atomic positions (SOAP), Localized graph kernel (LGK), Average kernel (AK), Kernel ridge regression (KRR), Gaussian process regression (GPR)}.}
\label{Tab::BA10-resume}
\end{table}
        
\section{Conclusions}        

In conclusion, we introduced in the framework of a local decomposition kernel a new similarity measurement based on molecular graphs. This kernel is composed by two parts: one that describes the local molecular pattern through a labeled graph, and a second one, that keeps into account some finer geometric information using Coulomb labels. These two kernels are hybridized through a hyperparameter $\alpha$ which controls their relative contributions. 

The accuracy of this new kernel was tested on two datasets: The molecular QM7 dataset and the BA10 dataset, containing ten binary alloys. The first of these applications involves the prediction of the atomization energies of molecules. In this case our method outperforms previous descriptors proposed in the literature such as, for example, the smooth overlap of atomic positions and a previous approach based on unlabeled graph kernel introduced by Ferré \textit{et al.} \cite{ferre2017}. In the case of enthalpies of formations of solids (BA10 dataset), for small training set size -- up to 300 datapoints -- the accuracy of our new HMPP kernel is lower with respect to the SOAP approach. However, when the number of training configurations increases the learning ability of the SOAP descriptor saturates. The HMPP kernel does not suffer of this limitation and outperforms SOAP for 1000 and more training configurations. From these case studies the labeled graph kernel seems a particularly promising tool to improve the accuracy in machine learning regression problems in chemistry and materials science.\\   

\section{Acknowledgement}

B.C. and D.R. acknowledge G. Ferr\'e for sharing his GRAPE code and for fruitful discussions. This work was supported
through the COMETE project (COnception in silico de Mat\'eriaux pour
l'EnvironnemenT et l'Energie) co-funded by the European Union under
the program ``FEDER-FSE Lorraine et Massif des Vosges 2014-2020''. 
T.B. acknowledges support from Slovak Research and
Development Agency under Contracts No. APVV-15-0105
and No. VEGA-1/0777/19.

\section{Data availability}

The data that support the findings of this study are available from the corresponding authors upon reasonable request.


\bibliography{Bibliographie-Kernels}

\end{document}